%% file: niisa.tex
\documentclass[11pt]{article} \usepackage[dnasubmission]{optional}
\usepackage{amsmath}
\usepackage{amssymb}
\usepackage{ifpdf}
\usepackage{algorithmic}
\usepackage[caption=false,format=hang]{subfig}
\usepackage{wrapfig}
\usepackage{cite}
\usepackage{graphicx}
\usepackage{enumitem}
\usepackage{tikz}   % draw pictures

\usepackage{commands-tam} % renamed \color (already defined) as \colorTAM
\usepackage{numbering-tam}
\usepackage{url}
\usepackage{amsfonts}
\usepackage{amsthm}
\usepackage{fullpage}

\opt{dnasubmission,dnafinal}{\newcommand{\figuresize}{\small}}
\opt{normal}{\newcommand{\figuresize}{\normalsize}}

\newcommand{\shrinkscale}{0.9}

\newcommand{\ignore}[1]{}

\ifpdf

  \usepackage[pdftex]{epsfig}
\else
    \usepackage[dvips]{epsfig}
    \newcommand{\href}[2]{#2}

\fi

\opt{els}{
    \theoremstyle{definition}
    \newtheorem{theorem}{Theorem}[section]
    
    \newtheorem{definition}[theorem]{Definition}

    \newenvironment{proof}{{\em Proof.~}}{\qed}

}
\begin{document}

\title{Negative Interactions in Irreversible Self-Assembly\thanks{This research was supported in part by Natural Sciences and Engineering Research Council of Canada (NSERC) Discovery Grant R2824A01 and the Canada Research Chair Award in Biocomputing to Lila Kari.}}
\date{}

\urldef\daveemail\url{ddoty@csd.uwo.ca}
\urldef\lilaemail\url{lila@csd.uwo.ca}
\urldef\benoitemail\url{benoit.masson@irisa.fr}

\author{David Doty\thanks{University of Western Ontario, Dept. of Computer Science, London, Ontario, Canada, \daveemail.}
\and
Lila Kari\thanks{University of Western Ontario, Dept. of Computer Science, London, Ontario, Canada, \lilaemail.}
\and
Beno\^{i}t Masson\thanks{IRISA (INRIA), Campus de Beaulieu, 35042 Rennes Cedex, France, \benoitemail.}
}
\maketitle

% ----------------------------------------------------------------

\input{0_abstract}
\input{1_intro}

\input{2_tam-informal}

\input{3_impossibility}
\input{4_tm}
\input{5_conclusion}

\begin{acks}
The authors are grateful to Shinnosuke Seki for insightful discussions.
\end{acks}

%\clearpage
\bibliographystyle{plain}
%{ \footnotesize \bibliography{main,dim,random,dimrelated,rbm,tam} }
{ \footnotesize \bibliography{tam} }

\opt{dnasubmission}{
    \clearpage
    \appendix
    \input{6_appendix}
}

\end{document}

%% file: 0_abstract.tex
\begin{abstract}
This paper explores the use of negative (i.e., repulsive) interactions in the abstract Tile Assembly Model defined by Winfree. Winfree in his Ph.D. thesis postulated negative interactions to be physically plausible, and Reif, Sahu, and Yin studied them in the context of \emph{reversible} attachment operations. We investigate the power of negative interactions with \emph{irreversible} attachments, and we achieve two main results. Our first result is an impossibility theorem: after $t$ steps of assembly, $\Omega(t)$ tiles will be forever bound to an assembly, unable to detach. Thus negative glue strengths do not afford unlimited power to reuse tiles. Our second result is a positive one: we construct a set of tiles that can simulate an $s$-space-bounded, $t$-time-bounded Turing machine, while ensuring that no intermediate assembly grows larger than $O(s)$, rather than $O(s \cdot t)$ as required by the standard Turing machine simulation with tiles.
\end{abstract}

%% text abstract
%This paper explores the use of negative (i.e., repulsive) interaction the abstract Tile Assembly Model defined by Winfree. Winfree postulated negative interactions to be physically plausible in his Ph.D. thesis, and Reif, Sahu, and Yin explored their power in the context of *reversible* attachment operations. We explore the power of negative interactions with *irreversible* attachments, and we achieve two main results. Our first result is an impossibility theorem: after t steps of assembly, Omega(t) tiles will be forever bound to an assembly, unable to detach. Thus negative glue strengths do not afford unlimited power to reuse tiles. Our second result is a positive one: we construct a set of tiles that can simulate a Turing machine with space bound s and time bound t, while ensuring that no intermediate assembly grows larger than O(s), rather than O(s * t) as required by the standard Turing machine simulation with tiles.

% keywords: self-assembly, self-destructible, negative bond strength, molecular computing, DNA computing

%% file: 1_intro.tex
\section{Introduction}
\label{sec-intro}

Tile-based self-assembly is a model of ``algorithmic crystal growth'' in which square ``tiles'' represent molecules that bind to each other via highly-specific bonds on their four sides, driven by random mixing in solution but constrained by the local binding rules of the tile bonds.
Erik Winfree \cite{Winf98}, based on experimental work of Seeman \cite{Seem82}, modified Wang's mathematical model of tiling
%\cite{Wang61,Wang63}
\cite{Wang61}
to add a physically plausible mechanism for growth through time.  Winfree defined a model of tile-based self-assembly known as the abstract Tile Assembly Model (aTAM). The fundamental components of this model are un-rotatable, but
translatable square ``tile types'' whose sides are labeled with ``glues'' representing binding sites. Two tiles that are placed next to each
other are attracted with strength determined by the glues where they abut, and in the aTAM, a tile \emph{binds} to an assembly if it is attracted on all sides with total strength at least a certain threshold value $\tau$.\footnote{The threshold $\tau$ models the temperature at which insufficiently strong chemical bonds will break, such as those formed by Watson-Crick complementarity in DNA-based implementations of tiles.} Assembly begins from a ``seed'' tile and progresses until no more tiles may attach.

We study a variant of this model in which glue strengths are allowed to be negative as well as positive. This leads to the situation in which a stable assembly may become unstable through the addition of a tile that, while binding strongly enough to the assembly to remain attached itself, exerts a repulsive force on a neighboring tile, which is sufficiently strong to detach some portion of the assembly. This is formally modeled by allowing an assembly to break into two parts any time that the two parts have total connection strength less than $\tau$ (i.e., if there is a cut of the interaction graph of strength less than $\tau$). Negative glue strengths were discussed as a plausible mechanism in Winfree's thesis \cite{Winf98}, and explored theoretically in a more general model of \emph{graph-based self-assembly} by Reif, Sahu and Yin \cite{ReifSahYin05}.
%\opt{normal}{
We compare the results of \cite{ReifSahYin05} to the present paper in more detail later in this section.
%}
%\opt{dnasubmission}{We compare the results of \cite{ReifSahYin05} to the present paper in more detail in Section \ref{sec-appendix}.}

This paper has two main contributions, an impossibility result and a positive result. The impossibility result is that under the irreversible model, negative glue strengths are not sufficient to achieve perfect reuse of tiles as in \cite{ReifSahYin05}.
    It is tempting to believe that with negative glue strengths, the monotonic growth of the aTAM could be overcome to such a degree that a bounded set of tiles could be reused for arbitrarily long computations,\footnote{subject, of course, to computational complexity constraints such as $\DTIME(t(n)) \subseteq \DSPACE(2^{t(n)})$, based on the observation that configurations cannot repeat during the course of a halting computation} hence implementing the observation that ``you can reuse space but you can't reuse time''.
    Alas, you cannot reuse space (tiles) too much with irreversible reactions. In this section, we show that under the irreversible model of tile assembly, even with negative glue strengths, many tiles will be forever bound to an assembly, unable to detach. In fact, this number is linear in the number of assembly operations, so that after $t$ operations, $\Omega(t)$ tiles will be permanently bound to some assembly.
%}
%\opt{normal}{\appA}

The positive result is a construction attempting to make do with this limitation.
    For concreteness, our construction shows how to simulate a single-tape Turing machine. But the idea applies to the iterated computation of any function $f$ that can be ``computed with constant height'' by a tile assembly system (a formal definition is given in Section \ref{sec-tm}). The function $f$ mapping the configuration of a Turing machine to its next configuration is an example of one such function. Other examples include the incrementing or decrementing of a counter, or the selection of a uniformly distributed random number from a finite set $\{1,2,\ldots,n\}$ using flips of a fair coin via von Neumann's \emph{rejection method}, as shown in \cite{RNSSA}.
%}
%\opt{normal}{\appB}

Our construction achieves the following property: if the Turing machine $M$ being simulated on input $x$ (with $n = |x|$) has space bound $s(n)$ and time bound $t(n)$, then $O(t(n) \cdot s(n))$ tiles (meaning total count of tiles, which is greater than the number of unique tile types), mixed in solution, will simulate the computation of $M$ on input $x$, and no intermediate assembly will grow to size larger than $O(s(n))$.
    The impossibility result can be interpreted to imply that external energy must be supplied to break bonds between tiles if we wish to reuse them for computation. If we wish to limit the volume of a solution, and therefore the number of molecules it can contain (by the finite density constraint, see \cite{SolCooWinBru08}) to $O(s(n))$, then we cannot allow intermediate assemblies to grow larger than this value. Of course, by the impossibility result, many more than $s(n)$ different such assemblies will form if $t(n) \gg s(n)$ (for instance, when simulating a logarithmic-space, polynomial-time computation). With a mechanism to ``vacuum'' away junk assemblies and supply the external energy needed to break them up (a mechanism not modeled in the aTAM), these tiles could be reused, bringing down the required number of tiles from $O(t(n) \cdot s(n))$ to $O(s(n))$.
    %}
%\opt{normal}{\appC}
%\opt{normal,dnasubmission}{Section \ref{subsec-model-choices} discusses this issue in greater detail.}

The main difference between \cite{ReifSahYin05} and the present paper is that \cite{ReifSahYin05} employs \emph{reversible} reactions, and the present paper employs \emph{irreversible} reactions.\footnote{\cite{ReifSahYin05} also uses a more general graph-based model of self-assembly, but this difference is less crucial than the reversibility issue.}
%\opt{dnasubmission}{Section \ref{sec-appendix} elaborates on the differences.}
    %\newcommand{\appD}{
    Within the aTAM, the main difference between our model and \cite{ReifSahYin05} amounts to a difference in the definition of a legal attachment operation. In~\cite{ReifSahYin05}, the authors define a tile attachment to be legal if the tile attaches with strength $\tau-1$ (in fact, they define it a bit differently but restricting attention to our construction and that of \cite{ReifSahYin05}, this definition is equivalent). This is a phenomenon not modeled by the aTAM, but it is physically plausible to suppose that it occurs, though with less frequency than strength $\tau$ attachments (see the kinetic TAM of \cite{Winf98}). Therefore the tile may detach after attaching since it is held with insufficient strength. But, if it first causes another tile or group of tiles to be bound with total strength less than $\tau$, then those tiles may also fall off, possibly resulting in stabilization of the original attachment. In the present paper, we define attachments to be legal only if they have strength at least $\tau$, whereas detachments may only happen between assemblies attached with strength at most $\tau-1$. This difference implies that our impossibility result does not apply to \cite{ReifSahYin05}, which can be considered an advantage of reversible interactions. But this advantage does not come without disadvantages: due to the second law of thermodynamics, their construction must necessarily be implemented as an unbiased random walk with equal rates of forward and reverse reaction, lest the entropy of the system increase with time if one direction is more favorable. Therefore their construction takes expected time $n^2$ to go forward $n$ steps.

    We should also note that although \cite{ReifSahYin05} uses a more general model of graph-based self-assembly, this does not imply that their construction of an assembly system simulating a space-bounded Turing machine simulation is a stronger result than our construction. The more general model affords more power to aid in a construction, such as allowing non-planar interactions, in addition to the extra power of reversible interactions. Therefore, we emphasize that our positive construction is not merely a specialization of the construction of \cite{ReifSahYin05} to grid graphs. The construction of \cite{ReifSahYin05} is inherently non-planar and reversible. A major source of the effort in designing our construction was getting it to work in the plane and use irreversible attachments. Many similar (and simpler) constructions that superficially appear to do the same thing as our construction do not actually work in our model, as they introduce not only a desired cut of strength less than $\tau$, but also some undesired cuts of strength less than $\tau$, which, if detached, will ruin the construction.
%}
%\opt{normal}{\appD}

\opt{normal}{
This paper is organized as follows.
\opt{normal}{Section \ref{sec-tam-informal} gives an description of the aTAM with negative glue strengths and discusses some issues associated with choosing a proper model of negative interactions in self-assembly.}
\opt{dnasubmission,dnafinal}{Section \ref{sec-tam-informal} gives an description of the aTAM with negative glue strengths.}
%\opt{normal,dnasubmission}{Section \ref{sec-tam} formally defines the aTAM.}
Section \ref{sec-impossibility} states and proves our impossibility result.
Section \ref{sec-tm} provides the construction for our positive result.
Section \ref{sec-conclusion} concludes the paper and discusses the utility of negative glue strengths in general.
}
%\opt{dnasubmission}{Section \ref{sec-appendix} is an appendix that discusses some issues associated with choosing a proper model of negative interactions in self-assembly and gives more details and discussion.}

For color figures, see \url{http://www.csd.uwo.ca/~ddoty/papers/niisa.pdf}.

%% file: 2_tam-informal.tex
\section{Abstract Tile Assembly Model}
\label{sec-tam-informal}

This section gives a brief definition of the abstract Tile Assembly Model (aTAM,~\cite{Winf98}) with negative glue strengths. This not a tutorial on the aTAM; for readers unfamiliar with the model, please see \cite{RotWin00} for an excellent introduction.
\opt{dnasubmission}{Section \ref{subsec-model-choices} discusses the options we identified in defining an appropriate model and explains the choices we made.}
%\opt{normal,dnasubmission}{The model is described formally, and more generally, in Section \ref{sec-tam}.}

\newcommand{\modelDiscussionSection}{
    \subsection{Issues with Choice of Model}
    \label{subsec-model-choices}

    There are many variations on the model of tile self-assembly with negative glue strengths. We identify six (somewhat) independent binary choices to be made:
    \begin{description}

        \item[seeded/unseeded:] In the seeded model, assembly starts from a specially designated seed tile. In the unseeded model, assembly may start with any tile.

        \item[single-tile addition/two-handed assembly:] In the single-tile addition model, tiles are added one at a time to an assembly. In two-handed assembly~\cite{AGKS05}, two assemblies, each possibly consisting of multiple tiles, may attach to each other.

        \item[irreversible/reversible:] In the irreversible model, for a tile/assembly to attach to an assembly, it must bind with strength $\geq \tau$ (though it may cause another cut of the resulting assembly to have strength $< \tau$). In the reversible model (used by Reif, Sahu, and Yin \cite{ReifSahYin05}), a tile/assembly may attach with strength $< \tau$, implying that it may detach (reverse), but may also cause another cut to detach. In \cite{ReifSahYin05}, most attachment events have the property that they cause precisely two cuts to have strength $\tau-1$, that of the attachment event itself, and another desired cut, and if the latter cut detaches, then the former cut now has strength $\tau$. Hence, on the assumption that each cut is equally likely, assembly proceeds in a unbiased random walk (thus taking expected time $n^2$ to go forward $n$ steps).

        \item[detachment precedes attachment/detachment and attachment in arbitrary order:] If detachment precedes attachment, this means we assume that whenever a negative-strength glue creates an unstable cut, every detachment event that can occur, will occur, until we are left with nothing but stable assemblies, before the next attachment event occurs. The more realistic model, detachment and attachment in arbitrary order, assumes that the operations of attach and detach are both legal at any stage. Therefore, if a negative-strength glue creates an unstable cut, \emph{but} this cut could be stabilized if an attachment secures it in place before it can detach, then we assume that this could happen.

        \item[finite tile counts/infinite tile counts:] Most models of tile self-assembly assume an infinite number of tile types, since a very large number can be easily created in a short time. However, if we wish to use negative glue strengths to implement space-bounded computation, and the finite density constraint (see \cite{SolCooWinBru08}) implies that the solution volume must be proportional to the number of molecules it contains, then we must constrain the number of molecules. See the next point for a discussion of another practical issue with using bounded space for computation with tile assembly.

        \item[tagged result/tagged junk:] This choice relates to how we designate what is the ``result'' of assembly. In the tagged result model, we designate a subset of tile types to be ``black'', and state that a result assembly is any terminal assembly with a black tile in it. This allows us to separate the junk from the result after assembly is complete, but does not allow junk to be removed \emph{during} assembly, since the black tile may not be attached to the result assembly until the very end of the assembly process.
            In the tagged junk model, we say that any assembly (terminal or not) with a black tile is junk and may be removed.\footnote{Actually it is more realistic to assume something like \emph{two} different black tiles, placed at a certain recognizable orientation (such as next to each other) to enforce that individual black tiles in solution are not interpreted as identifying junk until they actually attach to something, but we do not dwell on this issue since we use the tagged result model.}
            Furthermore, every producible assembly must have the property that it may either grow into the result, or will grow into an assembly tagged with a black tile as junk. That is, at any point in the assembly process, we could ``vacuum'' away all assemblies tagged with a black tile, knowing that anything remaining is not (yet) junk and can be re-used. This allows a computation with space bound $s$ and running time $t \gg s$ to be done in volume $O(s)$, so long as one uses detachment to ensure that no intermediate assembly grows larger than $O(s)$, by supplying just enough copies of tiles that are not swept away, and periodically sweeping away the tagged junk while supplying fresh unattached tiles to carry out more work.

            Note that neither of these tagging models are typically used in other tile self-assembly papers, but the special implications of negative glue strengths imply that we cannot simply follow the convention that the seed identifies the result, as discussed below.

    \end{description}

    Of these, the first three choices are incomparable in terms of power: each choice affords both advantages and disadvantages in terms of designing a tile assembly system. The latter three choices are more clear: for each choice, one option makes implementing a correct design strictly more difficult, but results in a more robust and realistically implementable construction. Respectively, these choices are 1) detachment and attachment in arbitrary order, 2) finite tile counts, and 3) tagged junk

    These choices are not completely independent. For instance, there are two (mathematical) uses for a seed tile: 1) to identify \emph{legal attachment events in single-tile addition}: attachment is legal if it is between a single tile and an assembly containing the seed, and 2) to identify the \emph{result of assembly with two-handed growth}: one must allow attachment of tiles separate from the seed, but identifies legal results as terminal assemblies containing the seed. We separate out these uses by making the method of ``tagging'' the result independent of the seed tile, using the seed only to define what assemblies are defined as producible. With negative glues implying the ability to remove the seed from an assembly, we must take care to allow attachment events in the seeded model not only with assemblies containing the seed, but with those \emph{derived from the seed}.

    For this extended abstract, we use the model of single-tile addition, irreversible, seeded, detachment and attachment in arbitrary order, infinite tile counts, and tagged result. We believe it is possible to modify our construction to allow unseeded, two-handed assembly, finite tile counts, and tagged junk (the latter two allowing a more realistic implementation of ``space-bounded computation'' as discussed above), but we delay this construction until the full version of this paper.
}

\opt{normal}{
  \modelDiscussionSection
  \subsection{Definition of Model}
}

$\Z$ and $\Z^+$ denote the set of integers and positive integers, respectively. Let $G$ be a finite alphabet of \emph{glues}. A \emph{tile type} is a tuple $t \in G^4$, i.e., a unit square with a glue on each side. Associated with the tile types is a \emph{glue strength function} $str:G\times G\to\Z$ that indicates, given two glues $g_1$ and $g_2$, the strength $str(g_1,g_2)$ with which they interact.
We assume a finite set $T$ of tile types, but an infinite number of copies of each tile type, each copy referred to as a \emph{tile}. Let $G(T)$ denote the set of all glues of tile types in $T$. An \emph{assembly}
(a.k.a., \emph{supertile})
is a positioning of tiles on the integer lattice $\Z^2$ (i.e., a partial function $\alpha:\Z^2 \dashrightarrow T$, where $\dashrightarrow$ denotes that the function is partial).
Each assembly induces a \emph{binding graph}, a grid graph whose vertices are tiles, with an edge between two tiles if they are adjacent (i.e., are Euclidean distance 1 apart).\footnote{Previous papers model the binding graph as having edges only between tiles that interact with positive strength. In the present paper, the presence of negative glue strengths means that we must consider every possible interaction between adjacent tiles, whether positive, negative, or 0.}
The assembly is \emph{$\tau$-stable}, or simply \emph{stable} if $\tau$ is understood from context, if every cut of its binding graph has weight (strength) at least $\tau$, where the weight of an edge is the strength of the glue it represents.
That is, the assembly is stable if at least energy $\tau$ is required to separate the assembly into two parts.
In this paper, where not stated otherwise, we assume that $\tau=2$.

A \emph{tile assembly system} (TAS) is a 4-tuple $\calT = (T,str,\sigma,\tau)$, where $T$ is a finite set of tile types, $str:G(T) \times G(T) \to \Z$ is the \emph{glue strength function}, $\sigma:\Z^2 \dashrightarrow T$ is the finite and $\tau$-stable \emph{seed assembly}, and $\tau\in\Z^+$ is the \emph{temperature}.
Given a TAS $\calT=(T,str,\sigma,\tau)$, an assembly $\alpha$ is \emph{producible} if either (base case) $\alpha = \sigma$, or (recursive case 1) $\alpha$ results from the $\tau$-stable attachment of a single tile to a producible assembly (``$\tau$-stable attachment'' meaning that the cut separating the tile from the rest of the assembly has strength $\geq \tau$), or (recursive case 2) $\alpha$ consists of one side of a cut of strength $< \tau$ of a producible assembly.
Note in particular that a producible assembly need not be stable, but may be stabilized by attachments before it can break apart.
An assembly $\alpha$ is \emph{terminal} if $\alpha$ is $\tau$-stable and no tile can be $\tau$-stably attached to $\alpha$.
Let $B \subseteq T$ be a set of ``black'' tile types.
$\calT$ is $B$-\emph{directed} (a.k.a., $B$-\emph{deterministic}, $B$-\emph{confluent}) if it has exactly one terminal, producible assembly containing one or more tiles from $B$.\footnote{We define this notion of $B$-directedness but do not henceforth discuss it explicitly, since our construction simulates a general ``computation'', and $B$ would depend on the goals of the computation being simulated. In our example construction in Section \ref{sec-tm} of simulating steps of a Turing machine, $B$ could, for instance, consist of the tile types that represent a halting state, so that only a terminal assembly representing the configuration of a halted Turing machine would be considered the result.}

\newcommand{\twoHandedModelDefn}{
    To define unseeded assembly, it suffices to drop $\sigma$ from the definition of TAS, and define the base case of a producible assembly as any individual tile. To define \emph{two-handed} assembly (a.k.a., \emph{multiple tile model}~\cite{AGKS05}), it suffices to change the first recursive case to state that legal attachment events are between any two producible assemblies, such that they can be positioned in such a way that the cut separating them has strength $\geq \tau$ (i.e., can be stably attached). Then, an assembly $\alpha$ is \emph{terminal} if for every producible assembly $\beta$, $\alpha$ and $\beta$ cannot be stably attached. 
    Figure~\ref{fig:2handed} illustrates the new behaviors allowed by the two-handed variant.

    \begin{figure}[!ht]
      \begin{center}
        \input{figures/2handed.tex}
      \end{center}
      \caption{\figuresize Typical example of two-handed assembly, at temperature $\tau = 2$. The segments between tiles represent the bonds, the number of segments encodes the strength of the bond (here,~$1$ or~$2$). In the seeded, single tile model with seed $\sigma = t_0$, the assembly at step (b) would be terminal.}
      \label{fig:2handed}
    \end{figure}
}

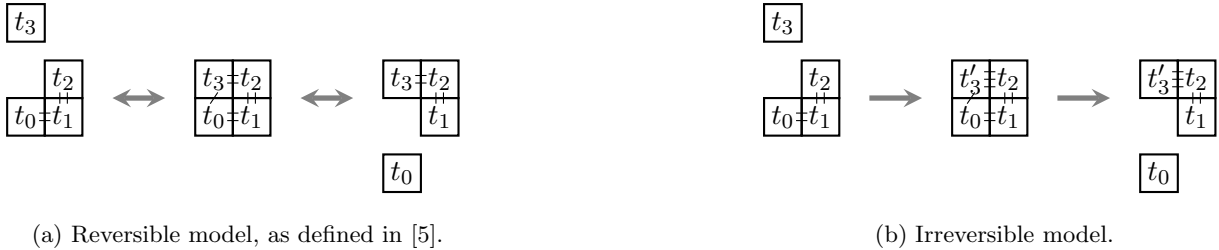
\begin{figure}[!ht]
      \begin{center}
      \subfloat[Reversible model, as defined in~\cite{ReifSahYin05}.]{\label{fig:neg1}
        \input{figures/negative1.tex}
      }\hfill
      \subfloat[Irreversible model.]{\label{fig:neg2}
        \input{figures/negative2.tex}
      }%\hfill\null
      \caption{\figuresize Two different implementations of negative interactions at temperature~$2$. The slanted bonds represent a strength of~$-1$. In the reversible model, the tile $t_3$ can attach with a total strength of~$1$ (one bond of strength~$2$ and one of strength~$-1$) and hence is unstable, while with our definition, $t'_3$ is attached with a total strength of~$2$ and forces $t_0$ to detach.}
      \label{fig:negative}
      \end{center}
    \end{figure}

    To define reversible assembly at temperature $\tau=2$ (as in \cite{ReifSahYin05}), it suffices to define attachment events with strength threshold $\tau-1 = 1$, rather than strength threshold $\tau=2$. This behavior is illustrated on Fig.~\ref{fig:negative}\subref{fig:neg1}, and can be compared with our new notion, whose evolution is shown on Fig.~\ref{fig:negative}\subref{fig:neg2}.

    \opt{dnasubmission,dnafinal}{\vspace{-30pt}}

%\opt{dnasubmission}{Section \ref{subsec-model-choices} discusses variations on the model that we have defined.}
\opt{normal}{
    To define unseeded assembly, it suffices to drop $\sigma$ from the definition of TAS, and define the base case of a producible assembly as any individual tile. To define \emph{two-handed} assembly (a.k.a., \emph{multiple tile model}~\cite{AGKS05}), it suffices to change the first recursive case to state that legal attachment events are between any two producible assemblies, such that they can be positioned in such a way that the cut separating them has strength $\geq \tau$ (i.e., can be stably attached). Then, an assembly $\alpha$ is \emph{terminal} if for every producible assembly $\beta$, $\alpha$ and $\beta$ cannot be stably attached. 
    Figure~\ref{fig:2handed} illustrates the new behaviors allowed by the two-handed variant.

    \begin{figure}[!ht]
      \begin{center}
        \input{figures/2handed.tex}
      \end{center}
      \caption{\figuresize Typical example of two-handed assembly, at temperature $\tau = 2$. The segments between tiles represent the bonds, the number of segments encodes the strength of the bond (here,~$1$ or~$2$). In the seeded, single tile model with seed $\sigma = t_0$, the assembly at step (b) would be terminal.}
      \label{fig:2handed}
    \end{figure}
}

%\opt{normal}{\otherModelDefns}
%comment out next line if we need space in DNA submission
%\opt{dnasubmission,dnafinal}{\otherModelDefns}

%% file: figures/2handed.tex
\begin{tikzpicture}[scale=0.06,>=stealth]
  \useasboundingbox (0,-5) rectangle (160,20);
  \draw[thick] (0,0) rectangle (10,10);
  \node at (5,5) {$t_0$};
  \draw (8.5,5) -- ++(3,0)
        (4,8.5) -- ++(0,3)
        (6,8.5) -- ++(0,3);
  \node at (5,-10) {\footnotesize (a)};
  \draw[->,color=gray,line width=2pt] (18,10) -- ++(14,0);

  \begin{scope}[shift={(40,0)}]
    \draw[thick] (0,0) rectangle (10,10);
    \node at (5,5) {$t_0$};
    \draw (8.5,5) -- ++(3,0)
          (4,8.5) -- ++(0,3)
          (6,8.5) -- ++(0,3);
    \draw[thick] (0,10) rectangle (10,20);
    \node at (5,15) {$t_1$};
    \draw (8.5,15) -- ++(3,0);
    \node at (5,-10) {\footnotesize  (b)};
    \draw[->,color=gray,line width=2pt] (18,10) -- ++(14,0);
  \end{scope}

  \begin{scope}[shift={(80,0)}]
    \draw[thick] (0,0) rectangle (10,10);
    \node at (5,5) {$t_0$};
    \draw (8.5,5) -- ++(3,0)
          (4,8.5) -- ++(0,3)
          (6,8.5) -- ++(0,3);
    \draw[thick] (0,10) rectangle (10,20);
    \node at (5,15) {$t_1$};
    \draw (8.5,15) -- ++(3,0);
    \node at (15,-10) {\footnotesize (c)};
    \begin{scope}[shift={(20,0)}]
      \draw[thick] (0,0) rectangle (10,10);
      \node at (5,5) {$t_2$};
      \draw (-1.5,5) -- ++(3,0)
            (4,8.5) -- ++(0,3)
            (6,8.5) -- ++(0,3);
      \draw[thick] (0,10) rectangle (10,20);
      \node at (5,15) {$t_3$};
      \draw (-1.5,15) -- ++(3,0);
      \draw[->,color=gray,line width=2pt] (18,10) -- ++(14,0);
    \end{scope}
  \end{scope}

  \begin{scope}[shift={(140,0)}]
    \draw[thick] (0,0) rectangle (10,10);
    \node at (5,5) {$t_0$};
    \draw (8.5,5) -- ++(3,0)
          (4,8.5) -- ++(0,3)
          (6,8.5) -- ++(0,3);
    \draw[thick] (0,10) rectangle (10,20);
    \node at (5,15) {$t_1$};
    \draw (8.5,15) -- ++(3,0);
    \node at (10,-10) {\footnotesize (d)};
    \begin{scope}[shift={(10,0)}]
      \draw[thick] (0,0) rectangle (10,10);
      \node at (5,5) {$t_2$};
      \draw (4,8.5) -- ++(0,3)
            (6,8.5) -- ++(0,3);
      \draw[thick] (0,10) rectangle (10,20);
      \node at (5,15) {$t_3$};
    \end{scope}
  \end{scope}

\end{tikzpicture}

%% file: figures/negative1.tex
\begin{tikzpicture}[scale=0.05,>=stealth]
  \draw[thick] (0,0) rectangle ++(10,10);
  \node at (5,5) {$t_0$};
  \draw[thick] (10,0) rectangle ++(10,10);
  \node at (15,5) {$t_1$};
  \draw[thick] (10,10) rectangle ++(10,10);
  \node at (15,15) {$t_2$};
  \draw[thick] (0,25) rectangle ++(10,10);
  \node at (5,30) {$t_3$};
  \draw (8.5,4) -- ++(3,0)
        (8.5,6) -- ++(3,0)
        (14,8.5) -- ++(0,3)
        (16,8.5) -- ++(0,3);
  \draw[<->,color=gray,line width=2pt] (28,10) -- ++(14,0);

  \begin{scope}[shift={(50,0)}]
    \draw[thick] (0,0) rectangle ++(10,10);
    \node at (5,5) {$t_0$};
    \draw[thick] (10,0) rectangle ++(10,10);
    \node at (15,5) {$t_1$};
    \draw[thick] (10,10) rectangle ++(10,10);
    \node at (15,15) {$t_2$};
    \draw[thick] (0,10) rectangle ++(10,10);
    \node at (5,15) {$t_3$};
    \draw (8.5,4) -- ++(3,0)
          (8.5,6) -- ++(3,0)
          (14,8.5) -- ++(0,3)
          (16,8.5) -- ++(0,3)
          (8.5,14) -- ++(3,0)
          (8.5,16) -- ++(3,0)
          (4,8.5) -- ++(2,3);
    \draw[<->,color=gray,line width=2pt] (28,10) -- ++(14,0);
  \end{scope}

  \begin{scope}[shift={(100,0)}]
    \path (0,-20);
    \draw[thick] (0,-15) rectangle ++(10,10);
    \node at (5,-10) {$t_0$};
    \draw[thick] (10,0) rectangle ++(10,10);
    \node at (15,5) {$t_1$};
    \draw[thick] (10,10) rectangle ++(10,10);
    \node at (15,15) {$t_2$};
    \draw[thick] (0,10) rectangle ++(10,10);
    \node at (5,15) {$t_3$};
    \draw (14,8.5) -- ++(0,3)
          (16,8.5) -- ++(0,3)
          (8.5,14) -- ++(3,0)
          (8.5,16) -- ++(3,0);
  \end{scope}
\end{tikzpicture}

%% file: figures/negative2.tex
\begin{tikzpicture}[scale=0.05,>=stealth]
  \path (0,-5);
  \draw[thick] (0,0) rectangle ++(10,10);
  \node at (5,5) {$t_0$};
  \draw[thick] (10,0) rectangle ++(10,10);
  \node at (15,5) {$t_1$};
  \draw[thick] (10,10) rectangle ++(10,10);
  \node at (15,15) {$t_2$};
  \draw[thick] (0,25) rectangle ++(10,10);
  \node at (5,30) {$t_3$};
  \draw (8.5,4) -- ++(3,0)
        (8.5,6) -- ++(3,0)
        (14,8.5) -- ++(0,3)
        (16,8.5) -- ++(0,3);
  \draw[->,color=gray,line width=2pt] (28,10) -- ++(14,0);

  \begin{scope}[shift={(50,0)}]
    \draw[thick] (0,0) rectangle ++(10,10);
    \node at (5,5) {$t_0$};
    \draw[thick] (10,0) rectangle ++(10,10);
    \node at (15,5) {$t_1$};
    \draw[thick] (10,10) rectangle ++(10,10);
    \node at (15,15) {$t_2$};
    \draw[thick] (0,10) rectangle ++(10,10);
    \node at (5,15) {$t'_3$};
    \draw (8.5,4) -- ++(3,0)
          (8.5,6) -- ++(3,0)
          (14,8.5) -- ++(0,3)
          (16,8.5) -- ++(0,3)
          (8.5,13) -- ++(3,0)
          (8.5,15) -- ++(3,0)
          (8.5,17) -- ++(3,0)
          (4,8.5) -- ++(2,3);
    \draw[->,color=gray,line width=2pt] (28,10) -- ++(14,0);
  \end{scope}

  \begin{scope}[shift={(100,0)}]
    \path (0,-20);
    \draw[thick] (0,-15) rectangle ++(10,10);
    \node at (5,-10) {$t_0$};
    \draw[thick] (10,0) rectangle ++(10,10);
    \node at (15,5) {$t_1$};
    \draw[thick] (10,10) rectangle ++(10,10);
    \node at (15,15) {$t_2$};
    \draw[thick] (0,10) rectangle ++(10,10);
    \node at (5,15) {$t'_3$};
    \draw (14,8.5) -- ++(0,3)
          (16,8.5) -- ++(0,3)
          (8.5,13) -- ++(3,0)
          (8.5,15) -- ++(3,0)
          (8.5,17) -- ++(3,0);
  \end{scope}
\end{tikzpicture}

%% file: 3_impossibility.tex
\section{Limitation of Tile Reuse with Irreversible Reactions}
\label{sec-impossibility}

If $\alpha$ is an assembly, define $\Phi(\alpha)$, the \emph{(negative) free energy} of $\alpha$, to be the sum of all glue strengths between adjacent tiles in the assembly.\footnote{The standard definition of free energy is the negative of this quantity, but as in \cite{RotWin00} we use its negation so that the quantity will be positive for stable assemblies. Intuitively, it is the energy \emph{required} to separate $\alpha$ into individual tiles, whereas the standard definition is the energy \emph{released} by such a separation.} In particular, an assembly consisting of a single tile has free energy 0.
If $S$ is a multiset of assemblies (such as that produced by a TAS with negative glue strengths, considering even the ``junk'' assemblies that are discarded after a cut), define the (negative) free energy of $S$ to be the sum of the free energies of each assembly in $S$, denoted $\Phi(S)$. Note that even postulating an infinite count of tiles, after a finite number of operations, only finitely many assemblies in $S$ consist of more than one tile, and each of these is a finite assembly. Therefore $\Phi(S) < \infty$ for any multiset $S$ of assemblies producible by a TAS, even in the case that $|S| = \infty$ (such as the initial multiset consisting of a countably infinite number of copies of each individual tile type).
%\opt{dnasubmission,normal}{(A formal definition is given in Section \ref{sec-free-energy}.)}

When we discuss the ``number of steps'' for the assembly process of a TAS, we mean the total number of attachment and detachment operations that have been applied so far. We do not claim that this is a proper model of ``running time'', but it is convenient to think of attachment and detachment events as discrete and equally-spaced steps, even though they may happen in parallel or with interval times governed by a continuous distribution.

\begin{theorem} \label{thm-impossibility}
Let $\calT$ be a TAS, and let $S$ be a multiset of assemblies producible by $\calT$ after $t\in\N$ steps. Then $\Phi(S) \geq t / 2$.
\end{theorem}

\newcommand{\impossibilityProof}{
\begin{proof}
%Suppose that $\calT$ starts with no tiles attached (i.e., $\calT$ is unseeded, or it is seeded with a seed of size 1);
Suppose that $\calT$ has a seed of size 1;
otherwise, the free energy we derive for step $t$ will be even higher, so this assumption does not harm the proof. For $i \in \{0,1,\ldots,t\}$, let $S_i$ denote the multiset of assemblies after the first $i$ operations, so that $S_t = S$ and $S_0$ is the multiset of individual unattached tiles. Note that $\Phi(S_0) = 0$. Let $A \subseteq \{1,2,\ldots,t\}$ be the indices of attachment operations, and let $D = \{1,2,\ldots,t\} - A$ be the indices of detachment operations, so that operation $i$ changes $S_{i-1}$ to $S_i$.

Each attachment operation increases the free energy by at least $\tau$ for a system operating at temperature $\tau$, since we require a tile/supertile attachment to have the property that the cut between it and the rest of the assembly has strength at least $\tau$, and the edges of this cut previously each contributed 0 to the free energy since they were all unbound. So for $S_{i-1}$ leading to $S_{i}$ via attachment, $\Phi(S_{i}) - \Phi(S_{i-1}) \geq \tau$. For each detachment operation, the greatest strength cut that could be broken to create the detachment has strength $\tau - 1$; stronger cuts cannot be broken. This implies the free energy decreases by at most $\tau - 1$ during a single detachment operation.\footnote{A single attachment of a tile with negative glue strength can potentially cause a cascade of detachments that, put together, lead to a large decrease in free energy. However, these are each considered separate detachment events.} So for $S_{i-1}$ leading to $S_{i}$ via detachment, $\Phi(S_{i}) - \Phi(S_{i-1}) \geq - (\tau - 1)$. Amortizing over all operations,\footnote{See \cite[Section 17.3]{CLRS01} for a discussion of amortized analysis, which is a fancy phrase for writing the following sum in this form.} we have that
\opt{normal}{
    \begin{eqnarray*}
        \Phi(S)
        &=&
        \sum_{i=1}^{t} (\Phi(S_i) - \Phi(S_{i-1}))
        \\&=&
        \sum_{i \in A} (\Phi(S_i) - \Phi(S_{i-1})) + \sum_{i \in D} (\Phi(S_i) - \Phi(S_{i-1}))
        \\&\geq&
        \sum_{i \in A} \tau + \sum_{i \in D} -(\tau - 1)
        \\&=&
        |A| \tau - |D|(\tau - 1).
    \end{eqnarray*}
}

\opt{dnasubmission,dnafinal}{
    $$
    \begin{array}{lclcl}
        \Phi(S)
        &=&
        \sum_{i=1}^{t} (\Phi(S_i) - \Phi(S_{i-1}))
        &=&
        \sum_{i \in A} (\Phi(S_i) - \Phi(S_{i-1})) + \sum_{i \in D} (\Phi(S_i) - \Phi(S_{i-1}))
        \\&\geq&
        \sum_{i \in A} \tau + \sum_{i \in D} -(\tau - 1)
        &=&
        |A| \tau - |D|(\tau - 1).
    \end{array}
    $$
}

Although we posit an infinite number of copies of each tile type, during the first $t$ steps at most $t+1$ tiles, denoted as the multiset $S'_0$, can actually participate in assembly operations.
Let $c_i$ denote the total number of assemblies in $S_i$ that consist of tiles from $S'_0$, so that $c_0$ is simply the total number of initial tiles that will participate in the first $t$ steps.
Each attachment event decreases $c_i$ by one, and each detachment event increases $c_i$ by one. Since $c_i \leq c_0$ for all $i$ (we cannot have more assemblies than there are tiles), this implies that for all $i$, the number of attachment events in the first $i$ steps is at least the number of detachment events in the first $i$ steps. Therefore $|A| \geq |D|$ and since $|A| + |D| = t$, we conclude $|A| \geq t/2$, whence the above inequality tells us that

\opt{normal}{
    \begin{align*}
        \Phi(S)
        &\geq
        |A| \tau - |D|(\tau - 1)
        \\&\geq
        |A| \tau - |A|(\tau - 1)
        \\&=
        |A|
        \\&\geq
        t / 2.\qedhere
    \end{align*}
}

\opt{dnasubmission,dnafinal}{
    \[
    \Phi(S)
    \geq
    |A| \tau - |D|(\tau - 1)
    \geq
    |A| \tau - |A|(\tau - 1)
    =
    |A|
    \geq
    t / 2.\qedhere
    \]
}
\end{proof}
}

\opt{normal}{\impossibilityProof

The proof works for two-handed systems, and since single-tile addition systems are simply two-handed systems with an extra constraint on legal attachment operations, the proof applies to single-tile addition systems as well. The proof also works both for seeded and unseeded systems. As there is no property of the 2D plane or grid graphs employed in our proof, the proof applies to the \emph{irreversible} version of the graph-based self-assembly model studied in a reversible context by Reif, Sahu, and Yin \cite{ReifSahYin05}.}
\opt{dnasubmission}{A proof of Theorem \ref{thm-impossibility} is given in Section \ref{sec:proof-impossibility}.}

In particular, since the glue strengths are bounded above by some constant $s$, this implies that after $t$ steps, at least $t / (2s)$ sides of tiles are bound. With the finite tile count assumption, once $t$ is sufficiently large that $t / (2s)$ exceeds the total number of sides available (i.e., 4 times the total number of tiles in solution), no more sides are available for binding, and self-assembly grinds to a halt. This is the sense in which a finite number of tiles cannot be reused indefinitely.

\begin{figure}[!ht]
  \begin{center}
    \input{figures/freefall.tex}
    \subfloat[Starting from the seed $\sigma$, $n-1$ tiles attach on the left, before a gray tile can be attached at the bottom.]{\hspace{0.29\textwidth}}\hfill
    \subfloat[From the gray tile, $n-1$ light gray tiles attach on the right, using the bonds on the top and on the left. The assembly also goes around these new tiles counter-clockwise, until $\sigma$ is reached.]{\hspace{0.29\textwidth}}\hfill
    \subfloat[The initial tiles are removed by negative bonds of strength $-2$, and replaced by $n$ new tiles until $t$ is attached. The next step is the detachment of the gray tile, followed by the detachment of the $n-1$ light gray tiles one after the other.]{\hspace{0.34\textwidth}}
    \caption{\figuresize A possible evolution where the stable addition of one tile (marked $t$) can lead to $n$ tiles detaching one after the other (black arrows), hence reducing the free energy by~$n$. As usual, the strength of bonds is represented by the number of segments between tiles, slanted bonds indicating a negative strength. Note that this is only one among many possible evolutions, since there may be several cuts of strength lower than $\tau$ that can be removed. In particular, at the last step, all the gray and light gray tiles can detach as one unique big cut, which will in turn break into pieces.}
    \label{fig:freefall}
  \end{center}
\end{figure}
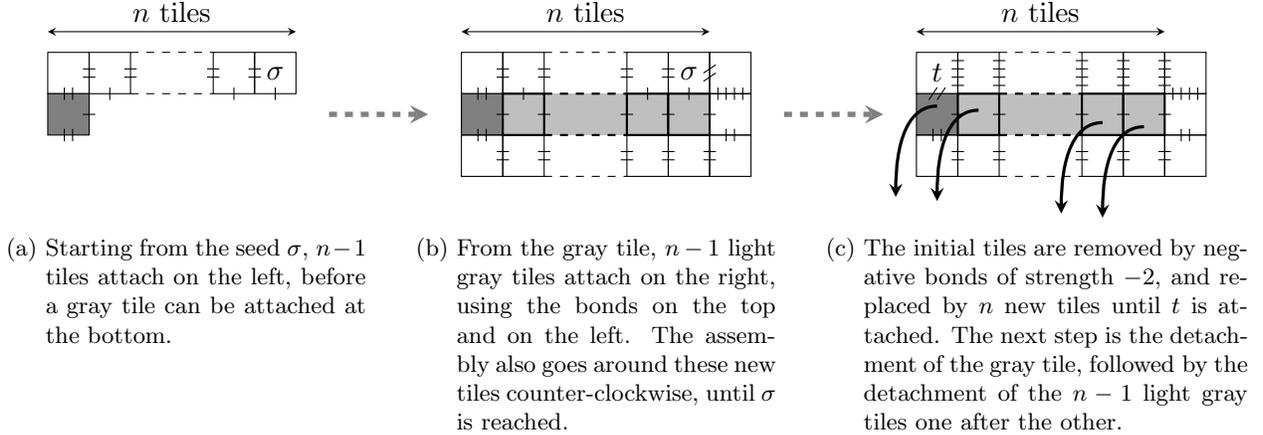

Some seemingly straightforward attempts to prove Theorem \ref{thm-impossibility} fail in ways that illustrate potentially nonintuitive properties of negative glue strengths.
It is not true, for instance, that the free energy increases monotonically, since it drops whenever a cut of positive strength is detached, so a straightforward inductive argument fails. Furthermore, it is not even true that the free energy decreases by at most a constant in between consecutive periods of increase. Even with fixed glue strengths (4 suffices), for each $n\in\N$, it is possible to construct a tile set with the property that there are two states of the system $S_i$ and $S_j$, with state $S_i$ preceding $S_j$, such that $\Phi(S_i) \geq \Phi(S_j) + n$. But, Theorem \ref{thm-impossibility} implies that any attempt to create such a cascade of detachments that drops the free energy by $n$ requires first attaching even stronger -- and ultimately unbreakable -- bonds required to set up the state $S_i$. That is, the free energy can fall arbitrarily far, but in order to do so it must first climb more than twice as high as it will eventually fall. This phenomenon is illustrated in Fig.~\ref{fig:freefall}, where the last stable addition of a tile leads to an arbitrary decrease of the free energy. This was made possible by the use of stronger strength-4 bonds prior to this event.

%\opt{dnasubmission,dnafinal}{\vspace{-30pt}}
%}

%\opt{normal}{\amortizedDiscussion}
%\amortizedDiscussion

There is a natural thermodynamic interpretation of Theorem \ref{thm-impossibility}: work done by tiles on tiles, in an irreversible manner, increases the entropy of the system by the second law of thermodynamics, thus decreasing the potential energy available to do more work. Therefore, any potential energy stored in the unattached glues is eventually permanently used up if external energy is not supplied to break these bonds. In our main construction, many junk assemblies are created that are no longer useful once the tiles in them have been used once. Theorem \ref{thm-impossibility} tells us that no amount of cleverness will allow us to break up those assemblies and reuse the tiles solely through design of tile types with negative glues; some external force must be supplied to break them apart using a mechanism not modeled in the aTAM.

Of course, Theorem \ref{thm-impossibility}, interpreted in light of the molecular interactions that are being modeled by the aTAM, should not be surprising to any physicist. But we believe it is important to formally establish the truth of such a statement within the model. One develops more confidence in a model of reality when it tells us something already known about reality (e.g., the Positive Mass Theorem \cite{SchYau79}).

Theorem \ref{thm-impossibility} does not apply to the negative glue strength construction of Reif, Sahu, and Yin \cite{ReifSahYin05}, because their model allows \emph{reversible} reactions.
\opt{normal}{Attempting to apply our proof to their model would result in the first inequality $\Phi(S_{i+1}) - \Phi(S_i) \geq \tau$ being replaced by $\Phi(S_{i+1}) - \Phi(S_i) \geq \tau - 1$, which would result in a final lower bound of 0, instead of $t/2$, for $\Phi(S)$.}
Intuitively, the reversibility of reactions implies that attachment and detachment have symmetric effects on the free energy. But this also implies that their system requires driving the system forward through an unbiased random walk, taking $n^2$ steps on average to proceed by $n$ net forward steps. Any attempt to speed up the reaction to make the forward rate of reaction faster than the reverse rate of reaction would introduce the imbalance in their respective effects on free energy that allows our proof to work. Therefore this tradeoff in speed versus reusability of tiles is fundamental.

%% file: figures/freefall.tex
\begin{tikzpicture}[scale=0.055,>=stealth]
  \useasboundingbox (0,-3) rectangle (280,40);
%  \draw[->,dash pattern=on 1.5pt off 1.5pt] (55,25) -| (5,13);
  \draw (0,20) rectangle ++(10,10);
  \draw (10,20) rectangle ++(10,10);
  \draw (40,20) rectangle ++(10,10);
  \draw (50,20) rectangle ++(10,10);
  \node at (55,25) {$\sigma$};
  \filldraw[fill=gray] (0,10) rectangle ++(10,10);
  \draw[dashed] (20,20) -- ++(20,0)
                (20,30) -- ++(20,0);
  \draw[<->] (0,35) -- node[above] {$n$ tiles} (60,35);
  \draw (8.5,24) -- ++(3,0)
        (8.5,26) -- ++(3,0)
        (18.5,24) -- ++(3,0)
        (18.5,26) -- ++(3,0)
        (38.5,24) -- ++(3,0)
        (38.5,26) -- ++(3,0)
        (48.5,24) -- ++(3,0)
        (48.5,26) -- ++(3,0)
        (4,18.5) -- ++(0,3)
        (6,18.5) -- ++(0,3)
        (4,8.5) -- ++(0,3)
        (6,8.5) -- ++(0,3);
  \draw (8.5,15) -- ++(3,0)
        (15,18.5) -- ++(0,3)
        (45,18.5) -- ++(0,3)
        (55,18.5) -- ++(0,3);
  \draw[->,color=gray,line width=2pt,dashed] (68,15) -- ++(24,0);

  \begin{scope}[shift={(100,0)}]
%    \draw[->,dash pattern=on 1.5pt off 1.5pt] (5,13.5) |- (65,5) -- (65,27);
    \draw (0,20) rectangle ++(10,10);
    \draw (10,20) rectangle ++(10,10);
    \draw (40,20) rectangle ++(10,10);
    \draw (50,20) rectangle ++(10,10);
    \node at (55,25) {$\sigma$};
    \filldraw[fill=gray] (0,10) rectangle ++(10,10);
    \draw (0,0) rectangle ++(10,10);
    \draw (10,0) rectangle ++(10,10);
    \draw (40,0) rectangle ++(10,10);
    \draw (50,0) rectangle ++(10,10);
    \draw (60,0) rectangle ++(10,10);
    \draw (60,10) rectangle ++(10,10);
    \draw (60,20) rectangle ++(10,10);
    \fill[fill=lightgray] (20,10) rectangle ++(20,10);
    \filldraw[fill=lightgray,thick] (10,10) rectangle ++(10,10);
    \filldraw[fill=lightgray,thick] (40,10) rectangle ++(10,10);
    \filldraw[fill=lightgray,thick] (50,10) rectangle ++(10,10);
    \draw[dashed] (20,20) -- ++(20,0)
                  (20,30) -- ++(20,0)
                  (20,0) -- ++(20,0);
    \draw[thick,dashed] (20,10) -- ++(20,0)
                        (20,20) -- ++(20,0);
%    \draw[->,dash pattern=on 1.5pt off 1.5pt] (6.5,15) -- (57,15);
    \draw[<->] (0,35) -- node[above] {$n$ tiles} (60,35);
    \draw (8.5,24) -- ++(3,0)
          (8.5,26) -- ++(3,0)
          (18.5,24) -- ++(3,0)
          (18.5,26) -- ++(3,0)
          (38.5,24) -- ++(3,0)
          (38.5,26) -- ++(3,0)
          (48.5,24) -- ++(3,0)
          (48.5,26) -- ++(3,0)
          (8.5,4) -- ++(3,0)
          (8.5,6) -- ++(3,0)
          (18.5,4) -- ++(3,0)
          (18.5,6) -- ++(3,0)
          (38.5,4) -- ++(3,0)
          (38.5,6) -- ++(3,0)
          (48.5,4) -- ++(3,0)
          (48.5,6) -- ++(3,0)
          (58.5,4) -- ++(3,0)
          (58.5,6) -- ++(3,0)
          (4,18.5) -- ++(0,3)
          (6,18.5) -- ++(0,3)
          (4,8.5) -- ++(0,3)
          (6,8.5) -- ++(0,3)
          (64,18.5) -- ++(0,3)
          (66,18.5) -- ++(0,3)
          (62,18.5) -- ++(0,3)
          (68,18.5) -- ++(0,3)
          (64,8.5) -- ++(0,3)
          (66,8.5) -- ++(0,3);
    \draw (8.5,15) -- ++(3,0)
          (18.5,15) -- ++(3,0)
          (38.5,15) -- ++(3,0)
          (48.5,15) -- ++(3,0)
          (58.5,15) -- ++(3,0)
          (15,18.5) -- ++(0,3)
          (45,18.5) -- ++(0,3)
          (55,18.5) -- ++(0,3);
    \draw (58.5,25) -- ++ (3,2)
          (58.5,23) -- ++ (3,2);
    \draw[->,color=gray,line width=2pt,dashed] (78,15) -- ++(24,0);
  \end{scope}

  \begin{scope}[shift={(210,0)}]
%    \draw[->,dashed] (65,25) -- (3,25);
    \draw (0,20) rectangle ++(10,10);
    \node at (5,25) {$t$};
    \draw (10,20) rectangle ++(10,10);
    \draw (40,20) rectangle ++(10,10);
    \draw (50,20) rectangle ++(10,10);
    \filldraw[fill=gray] (0,10) rectangle ++(10,10);
    \draw (0,0) rectangle ++(10,10);
    \draw (10,0) rectangle ++(10,10);
    \draw (40,0) rectangle ++(10,10);
    \draw (50,0) rectangle ++(10,10);
    \draw (60,0) rectangle ++(10,10);
    \draw (60,10) rectangle ++(10,10);
    \draw (60,20) rectangle ++(10,10);
    \fill[fill=lightgray] (20,10) rectangle ++(20,10);
    \filldraw[fill=lightgray,thick] (10,10) rectangle ++(10,10);
    \filldraw[fill=lightgray,thick] (40,10) rectangle ++(10,10);
    \filldraw[fill=lightgray,thick] (50,10) rectangle ++(10,10);
    \draw[dashed] (20,20) -- ++(20,0)
                  (20,30) -- ++(20,0)
                  (20,0) -- ++(20,0);
    \draw[thick,dashed] (20,10) -- ++(20,0)
                        (20,20) -- ++(20,0);
    \draw[<->] (0,35) -- node[above] {$n$ tiles} (60,35);
    \draw (8.5,24) -- ++(3,0)
          (8.5,26) -- ++(3,0)
          (8.5,22) -- ++(3,0)
          (8.5,28) -- ++(3,0)
          (18.5,24) -- ++(3,0)
          (18.5,26) -- ++(3,0)
          (18.5,22) -- ++(3,0)
          (18.5,28) -- ++(3,0)
          (38.5,24) -- ++(3,0)
          (38.5,26) -- ++(3,0)
          (38.5,22) -- ++(3,0)
          (38.5,28) -- ++(3,0)
          (48.5,24) -- ++(3,0)
          (48.5,26) -- ++(3,0)
          (48.5,22) -- ++(3,0)
          (48.5,28) -- ++(3,0)
          (58.5,24) -- ++(3,0)
          (58.5,26) -- ++(3,0)
          (58.5,22) -- ++(3,0)
          (58.5,28) -- ++(3,0)
          (8.5,4) -- ++(3,0)
          (8.5,6) -- ++(3,0)
          (18.5,4) -- ++(3,0)
          (18.5,6) -- ++(3,0)
          (38.5,4) -- ++(3,0)
          (38.5,6) -- ++(3,0)
          (48.5,4) -- ++(3,0)
          (48.5,6) -- ++(3,0)
          (58.5,4) -- ++(3,0)
          (58.5,6) -- ++(3,0)
          (4,8.5) -- ++(0,3)
          (6,8.5) -- ++(0,3)
          (64,18.5) -- ++(0,3)
          (66,18.5) -- ++(0,3)
          (62,18.5) -- ++(0,3)
          (68,18.5) -- ++(0,3)
          (64,8.5) -- ++(0,3)
          (66,8.5) -- ++(0,3);
    \draw (8.5,15) -- ++(3,0)
          (18.5,15) -- ++(3,0)
          (38.5,15) -- ++(3,0)
          (48.5,15) -- ++(3,0)
          (58.5,15) -- ++(3,0);
    \draw (3,18.5) -- ++ (2,3)
          (5,18.5) -- ++ (2,3);
    \foreach \x/\y in {0/1,10/0,40/-3,50/-4} {
      \begin{scope}[shift={(\x,\y)}]
        \draw[very thick,->] (5,16) .. controls ++(-10,0) and ++(0,5) .. ++(-10,-22);
%        \draw[->,line width=2pt] (5,16) -- (-3,0) -- (4,-1) -- (0,-9);
%        \draw[->,color=white,line width=1pt] (4.8,15.6) -- (-3,0) -- (4,-1) -- (0.3,-8.4);
      \end{scope}
    }
  \end{scope}
\end{tikzpicture}

%% file: 4_tm.tex
\section{Turing Machine Simulation}
\label{sec-tm}

Throughout this section, fix some finite alphabet $\Sigma$.
We first describe the class of functions that we will compute, which are intuitively those computable by a constant number of rows of assembly (although the number of columns is unbounded) in the standard aTAM. See \cite{RNSSA}, for example, for a formal definition of the standard aTAM model. Briefly, it is the same as the model defined in Section \ref{sec-tam-informal}, but glue strengths are non-negative and are only positive between equal glues.

\newcommand{\constantRowDefn}{
    \begin{definition}
    Let $T$ be a set of tile types, and let $e:T \to \Sigma$. We say that a row of tiles (a connected subassembly of some assembly of height 1) $t_1,t_2,\ldots,t_k$ \emph{$e$-encodes} a string $x\in\Sigma^k$ if $e(t_1)=x[1],e(t_2)=x[2],\ldots,e(t_k)=x[k]$, where $x[i]\in\Sigma$ is the $i^\text{th}$ symbol in $x$. A function $f:\Sigma^* \to \Sigma^*$ is \emph{constant-row computable} if there exists a tile set $T$, a function $e:T \to \Sigma$, and a constant $c$ such that, for each $x \in \Sigma^*$, there is a height-1 stable assembly $\sigma_x:\Z^2 \dashrightarrow T$ $e$-encoding $x$ such that the tile assembly system $\calT = (T,str,\sigma_x,2)$ (with $str(g_1,g_2) > 0 \iff g_1=g_2$) has the unique terminal assembly $\alpha$, the height of $\alpha$ is $c$, the bottom row of $\alpha$ is $\sigma_x$, the top row of $\alpha$ $e$-encodes $f(x)$, and the leftmost column of any row of $\alpha$ is no further left than the bottom row.
    \end{definition}
}
\opt{normal}{\constantRowDefn}
\opt{dnasubmission,dnafinal}{
    Informally (see Section~\ref{sec:def-CRC} for a formal definition), a function $f:\Sigma^* \to \Sigma^*$ is \emph{constant-row computable} if there is a tile set $T$ and a constant $c\in\N$ such that, for each $x\in\Sigma^*$, if a horizontal row $\sigma_x$ of tiles from $T$ that ``represents'' $x$ in a straightforward way is used as the seed assembly, then the resulting tile assembly system, at temperature 2, will grow into an assembly whose top row ``represents'' $f(x)$. Furthermore, the entire assembly will have height exactly $c$, and the left most tile of each row will have the same horizontal position as the left-most tile of the bottom row.
}

The widths of the rows representing the input and output may be different (i.e., possibly $|x| \neq |f(x)|$). In this case, we require only that the leftmost and rightmost tiles of each row have their glues specially marked to distinguish them from the tiles interior to the row.

Our construction shows how to design a tile set that will compute iterations of any constant-row computable function $f$, ensuring that no intermediate assembly grows larger than the size of the input or output processed by any \emph{individual} invocation of $f$. Examples of such functions include the function $f$ that, given a configuration of a single-tape Turing machine outputs the next configuration of this Turing machine, or %the function
 that increments a counter represented in binary.
%For concreteness, we use the example of the simulation of a single-tape Turing machine with space bound $s$ and time bound $t$ (possibly with $t \gg s$), while guaranteeing that no intermediate assembly grows to size larger than $O(s)$.

\newcommand{\figureonecaption}{High-level overview of assembly for computation of a constant-row computable function $f$.\ignore{ First the scaffold tiles connect to the $x$ data assembly. The scaffold tiles initiate the computation of $f$, and ``detect'' when the computation is finished, in the sense that the green row above $f(x)$ tiles cannot complete until all of $f(x)$ is present. Then the scaffold tiles grow back to the first scaffold tile to initiate the removal of $f(x)$ from the tiles surrounding $f(x)$. The removal tiles each use a negative glue strength against the tile ``in front of'' (on the path show by the arrows) it, and once this tile is removed, a new removal tile grows in its place to continue the removal. The path and bond placements and strengths are carefully chosen to ensure that no portion of $f(x)$ is removed, until the last step when $f(x)$ detaches whole from the rest of the tiles.
The width of the remaining ``junk'' assembly is a constant plus $O(1) + \max\{|x|,|f(x)|\}$, and the height is constant since $f$ is constant-row computable, so the size of the intermediate assembly is $O(\max\{|x|,|f(x)|\})$.}
}

\begin{figure}
\begin{center}
  \opt{normal}{\includegraphics[width=6.5in]{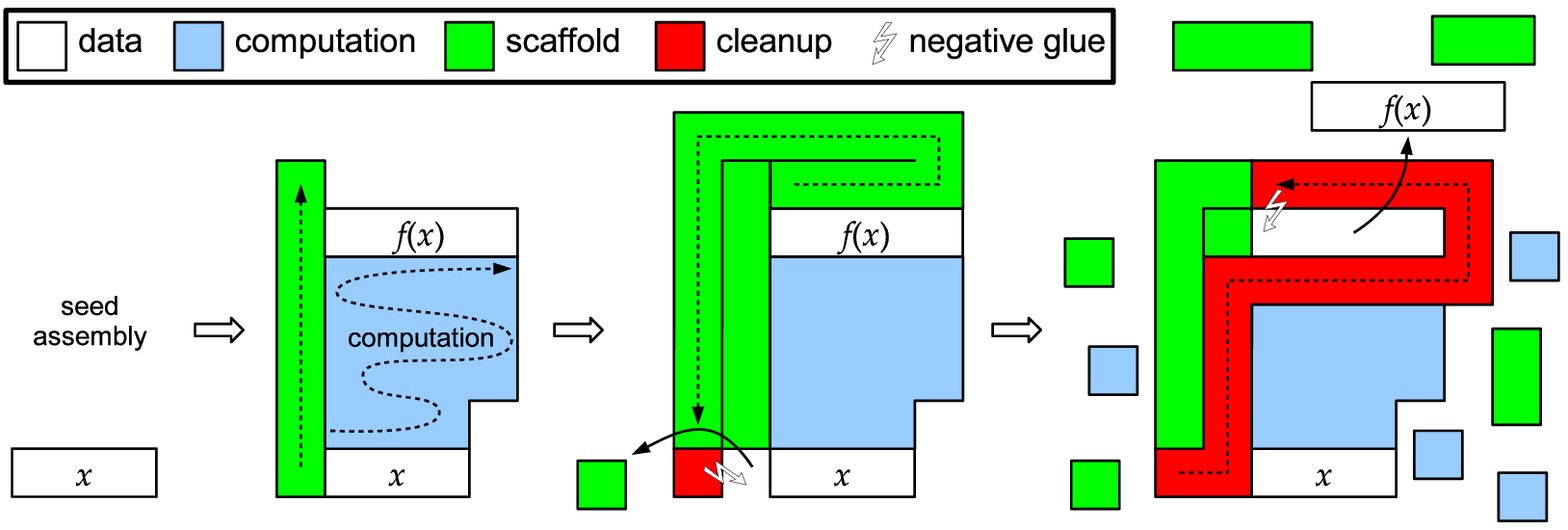}}
  \opt{dnasubmission,dnafinal}{\includegraphics[scale=\shrinkscale]{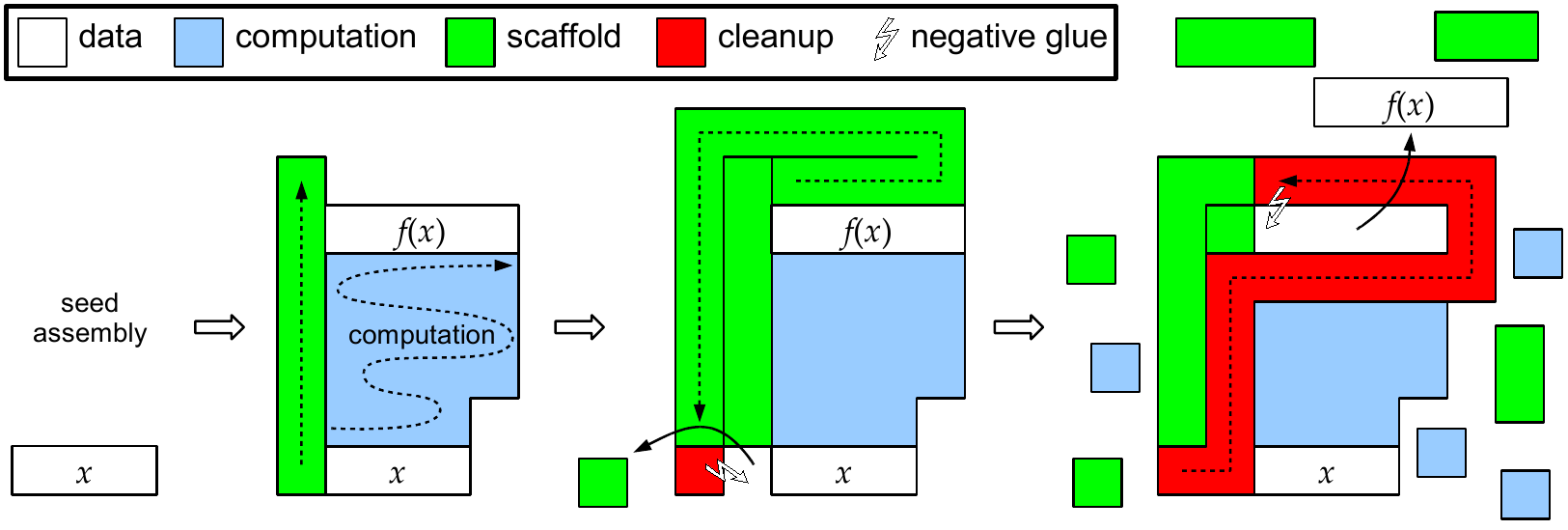}
    \par
    {\footnotesize \hspace{1.3cm} \phantom{(a)} \hspace{2.35cm} (a) \hspace{3.6cm} (b) \hspace{3.9cm} (c) \hfill}}
  \caption{\label{fig:high-level-overview} \figuresize \figureonecaption }
\end{center}
\end{figure}%\opt{dnasubmission,dnafinal}{\vspace{-30pt}}

Figure \ref{fig:high-level-overview} shows a high-level overview of the entire construction, in terms of a general constant-row computable function $f$. For concreteness, think of $f$ as the function that, given a configuration of a $t$-time-bounded, $s$-space-bounded, single-tape Turing machine, outputs the next configuration of this Turing machine (extending the tape on the right side only). The construction proceeds as follows, each label corresponds to a picture in Figure~\ref{fig:high-level-overview}.
\begin{enumerate}[label=(\alph*),itemsep=0pt,topsep=1ex]
%  \item A seed block encodes the initial data $x$.
  \item First, the scaffold tiles (green) connect to the $x$ data assembly (white). The scaffold tiles initiate the computation of $f$ (blue).
  \item The scaffold ``detects'' when the computation is finished, in the sense that the green row above $f(x)$ tiles cannot complete until all of $f(x)$ is present. Then the scaffold tiles grow back to the first scaffold tile to initiate the removal of $f(x)$ from the tiles surrounding $f(x)$.
  \item The removal tiles (red) each use a negative glue strength against the tile ``in front of'' (on the path show by the arrows) it, and once this tile is removed, a new removal tile grows in its place to continue the removal. The path and bond placements and strengths are carefully chosen to ensure that no portion of $f(x)$ is removed, until the last step when $f(x)$ detaches whole from the rest of the tiles.
\end{enumerate}

Note that since $f$ is constant-row computable, the height of the scaffold and removal parts are bounded by a constant and therefore may be hard-coded into the tile set, whereas special glues mark the horizontal endpoints so that the length of $x$ and $f(x)$ are not constrained.

The simulation of the Turing machine for $t$ steps will then consist of executing this assembly process %described in Figure \ref{fig:high-level-overview}
 for $t$ iterations, using the output assembly $f(x)$ as the input assembly $x$ for the next execution.
%The function $f$ is left unspecified, as the construction works for other functions such as incrementing or decrementing of a binary counter, as discussed in the introduction.
After each iteration, the width of the remaining ``junk'' assembly is a constant plus $O(1) + \max\{|x|,|f(x)|\}$, and the height is constant since $f$ is constant-row computable, so the size of the intermediate assemblies is $O(\max\{|x|,|f(x)|\})$.

%Scaffold tiles initiate the computation when they attach to $x$, ensuring that this computation is complete before the cleanup tiles remove the scaffold, computation, and $x$ tiles from the $f(x)$ tiles.

\medskip

\opt{normal,dnasubmission}{
    Figures \ref{fig:computation}, \ref{fig:position-for-cleanup}, and \ref{fig:cleanup}
    %(Appendix~\ref{subsec-main-construction-details})
    %respectively illustrate the action between the block arrows in Figure \ref{fig:high-level-overview}
    give some more details for the three main steps of Figure \ref{fig:high-level-overview}, respectively (a), (b), and (c)%
    , using the specific example of $f$ mapping a configuration of a single-tape Turing machine to its next configuration.
}
\opt{dnafinal}{
    We omit the details of the construction in this extended abstract.
}

\newcommand{\mainConstructionDetails}{
    \newcommand{\figuretwocaption}{Example of tiles implementing the computation step. Arrows within tiles show order of growth. In this case $f$ is constant-row computable with constant $c=1$. The first and last copy rows, shown in lighter shade than the center computation tiles, are always present no matter the function $f$, and their placement is initiated by the scaffold tiles. However, there is no interaction between the center computation and scaffold tiles. Note that the data tiles are two rows with strength 1 glues; this is to make them stable at temperature 2 but not producible (without additional scaffolding) as they would be if they were a single row connected with strength 2 glues.}

    \newcommand{\figurethreecaption}{Tiles that position the cleanup tiles. Here the ``copy'' tiles from Figure \ref{fig:computation} are depicted in the same shade as the computation tiles; now that $f(x)$ has been computed our goal is to remove all of them from the subassembly representing $f(x)$. The order of growth of the scaffold tiles ensures that cleanup does not begin until all of $f(x)$ is present.}

    \begin{figure}
    \begin{center}
      \opt{normal}{\includegraphics[width=6.5in]{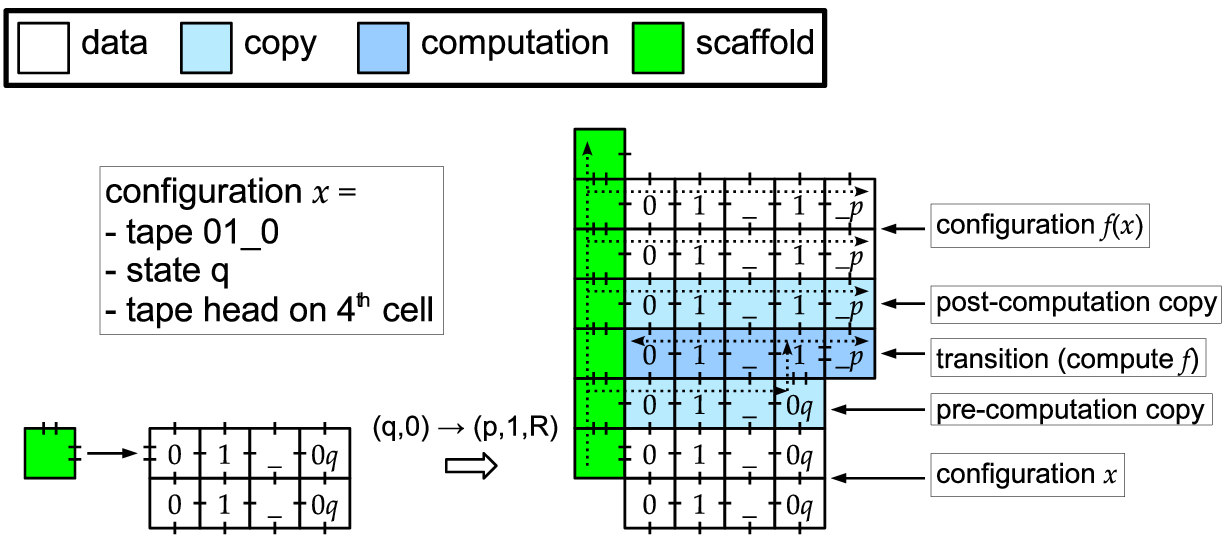}}
      \opt{dnasubmission,dnafinal}{\includegraphics[scale=\shrinkscale]{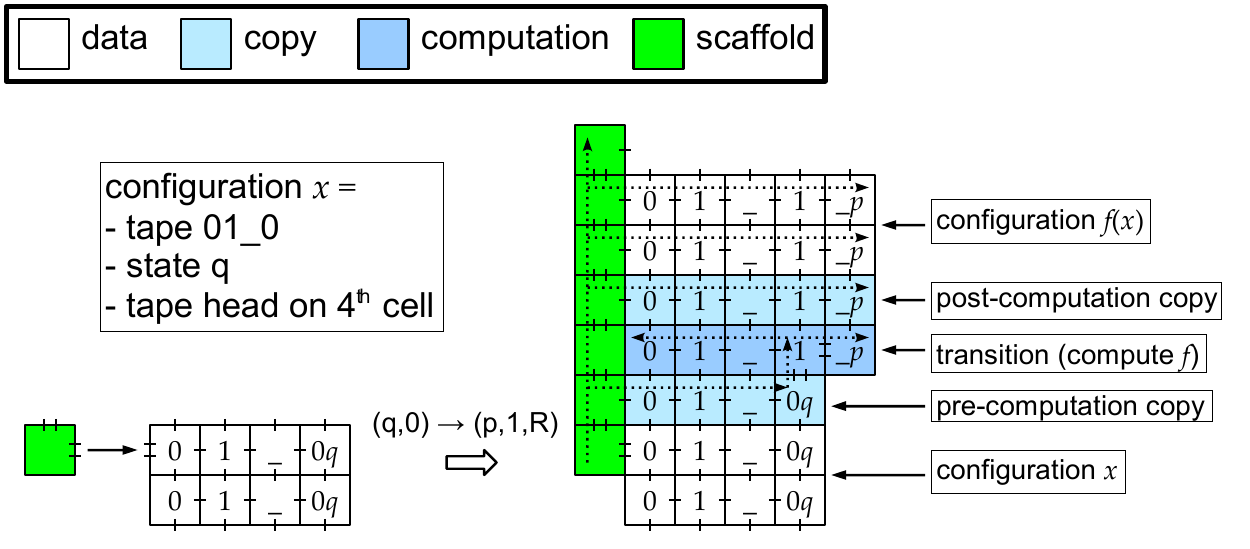}}\\
      \caption{\label{fig:computation} \figuresize \figuretwocaption }
    \end{center}
    \end{figure}

    Figure \ref{fig:computation} shows an example of tiles implementing step (a) of Figure \ref{fig:high-level-overview}, i.e., the computation of $f$.  The example shows one transition of a single-tape Turing machine, with tape contents $01\_0$ ($\_$ standing for blank), in state $q$, with tape head on the rightmost cell, transitioning to state $p$, moving the tape head right, changing the cell's symbol from 0 to 1, and encountering a blank on the new rightmost cell. In this case, a new rightmost cell is needed, illustrating how our construction handles dynamically changing space requirements, but if the tape head were further left in the row, it would simply fill in copy tiles to the right, just as to the left as shown above, and the row would stay the same width.
    At the start and end of a computation, the configuration is copied so that any strength $> 1$ bonds used in the computation are on the interior of the computation tiles, ensuring that only strength-1 bonds must later be broken to separate the data tiles. Each data assembly on either end of the computation tiles is represented by a two-row assembly with only single-strength bonds on its interior, which ensures that when detached, the data assembly will be stable, but that it cannot form on its own without help from the scaffold tiles (which would happen if it were only a single row connected with strength-2 bonds). Each vertical position is hard-coded into the tile set; i.e., the scaffold tile set ``knows'' the required height to compute $f$. However, the absolute horizontal positions are not encoded into the tiles, only the leftmost and rightmost tiles of the configuration are specially marked, and all interior tiles representing the same data are identical.

    \begin{figure}
    \begin{center}
      \opt{normal}{\includegraphics[width=6.5in]{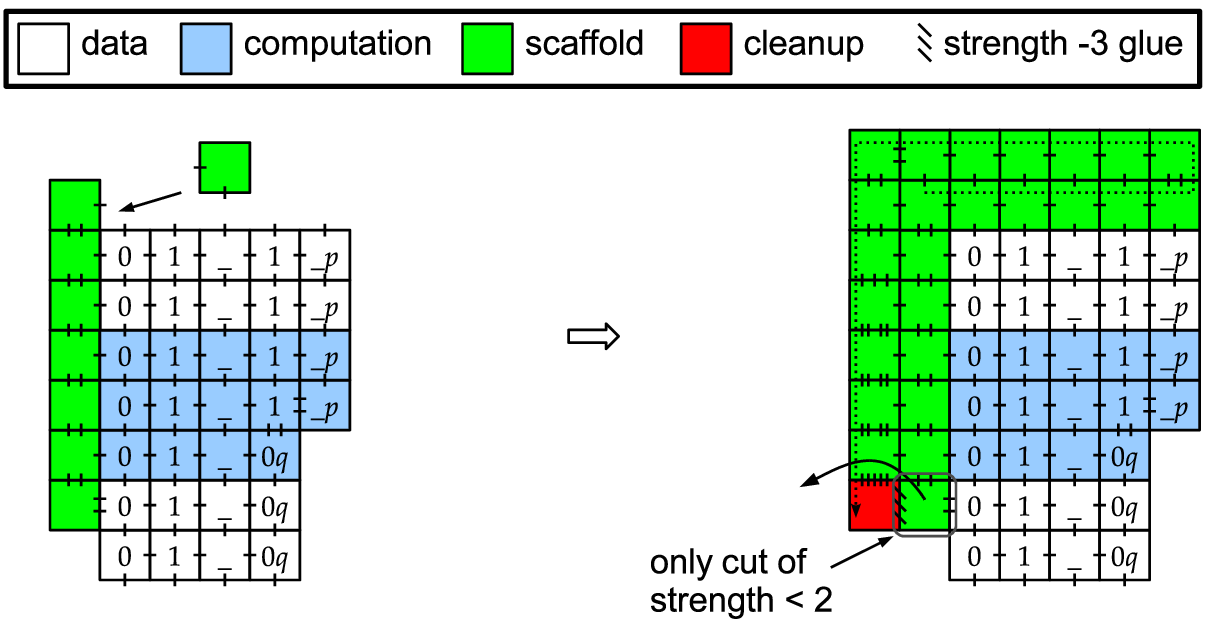}}
      \opt{dnasubmission,dnafinal}{\includegraphics[scale=\shrinkscale]{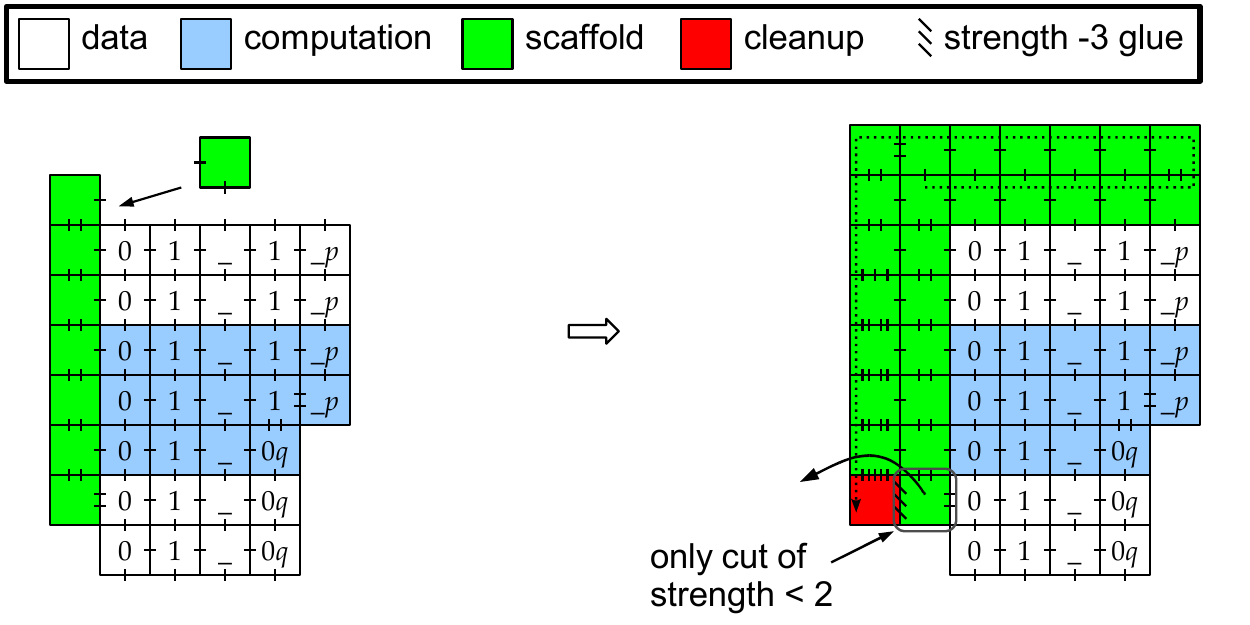}}
      \caption{\label{fig:position-for-cleanup} \figuresize \figurethreecaption }
    \end{center}
    \end{figure}

    Figure \ref{fig:position-for-cleanup} shows the tiles implementing step (b) of Figure \ref{fig:high-level-overview}, positioning the tiles for cleanup.
    The top two rows must use cooperation to tell where the end of the row underneath is, since the width of the output row is unknown. The strengths of bonds on the leftmost downward-growing column must be sufficiently large to ensure that only the proper cut is made when the first negative-strength glue is applied.

    \newcommand{\figurefourcaption}{
      Tiles that ``clean up'' the connections between the output data and the scaffold and computation tiles to separate them and allow the data tiles to be computed on again.
    }

    \begin{figure}
    \begin{center}
      \opt{normal}{\includegraphics[width=6.5in]{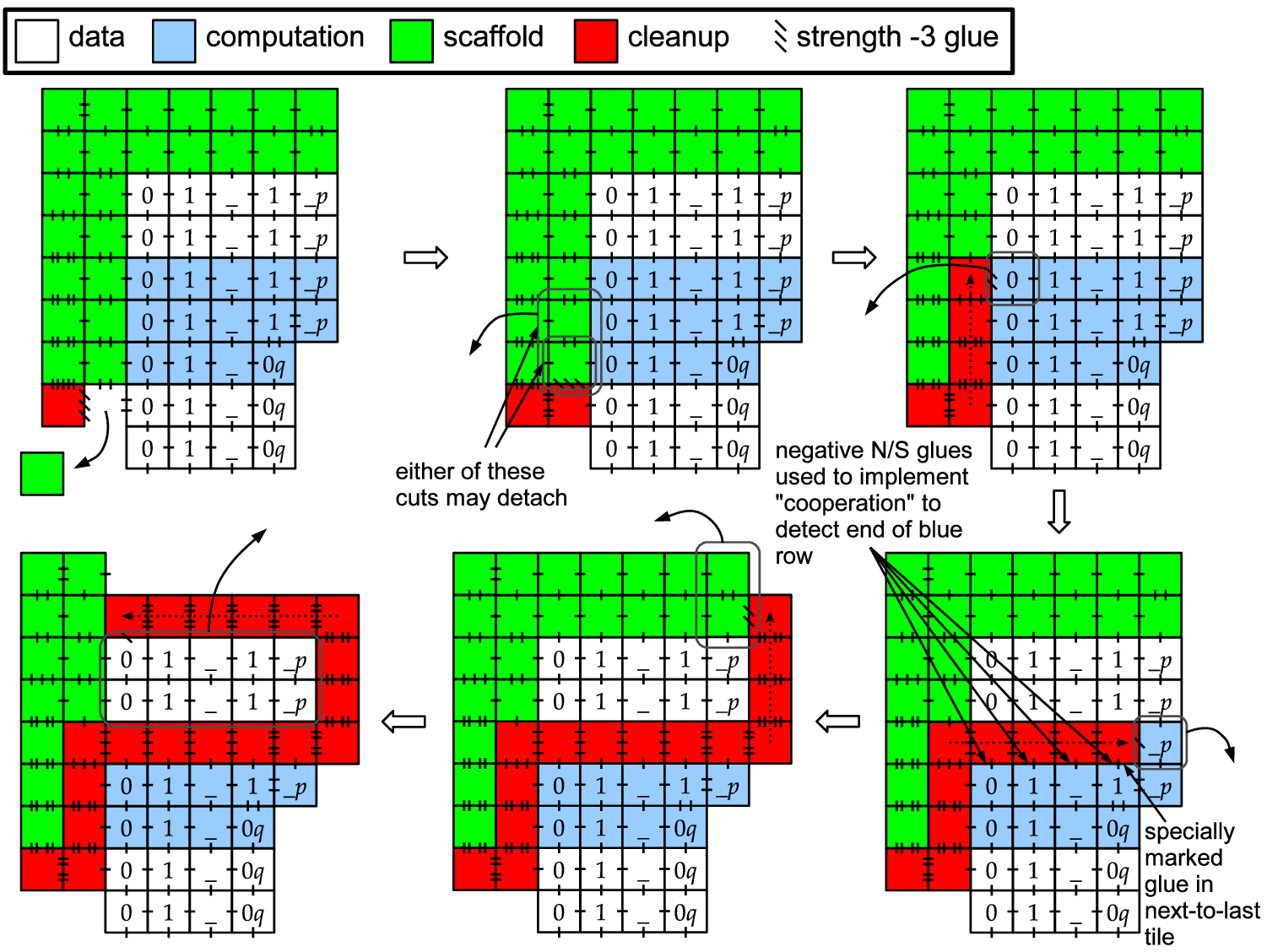}}
      \opt{dnasubmission,dnafinal}{\includegraphics[scale=\shrinkscale]{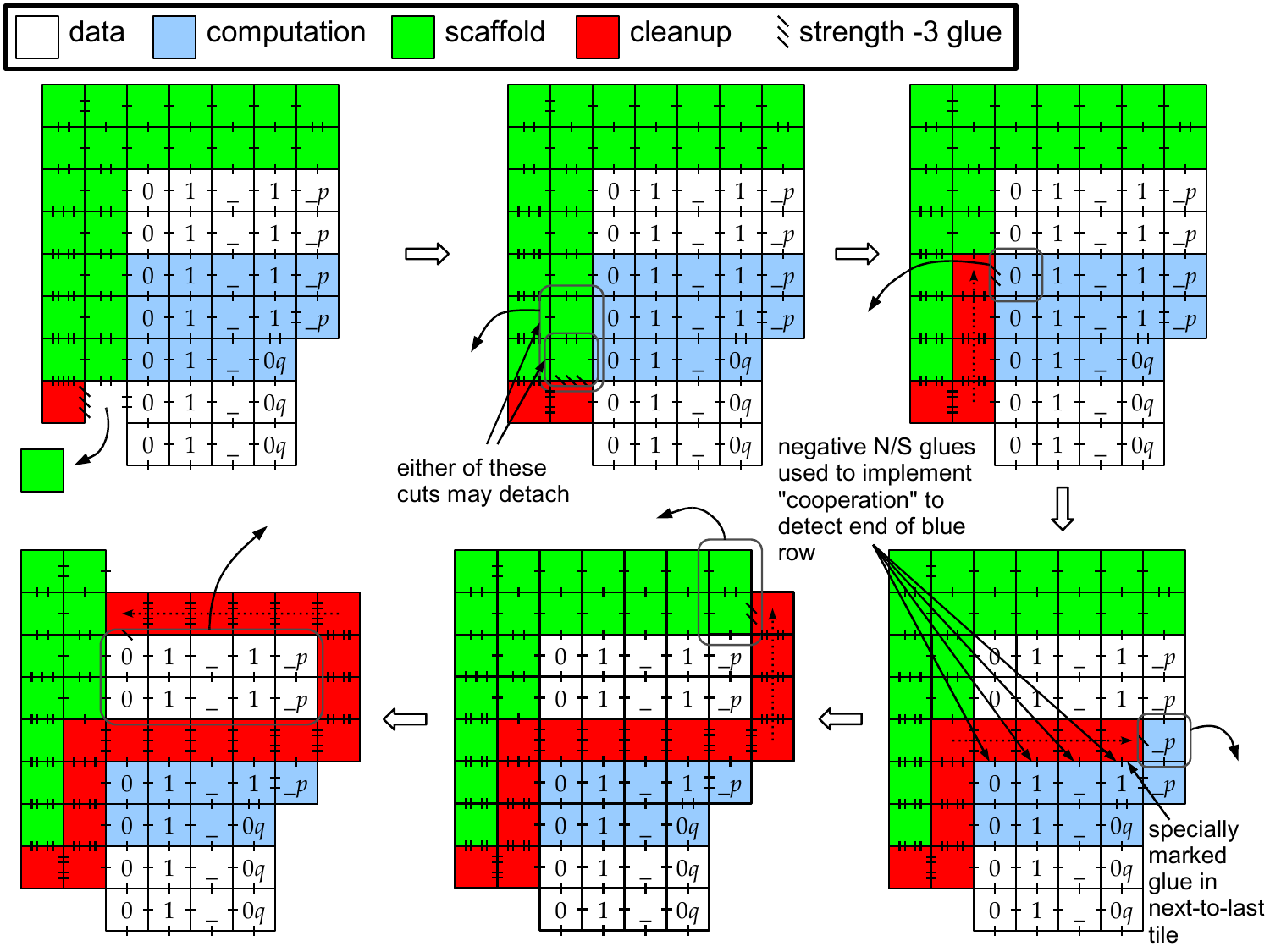}}
      \caption{\label{fig:cleanup} \figuresize \figurefourcaption}
    \end{center}
    \end{figure}

    Figure \ref{fig:cleanup} shows the tiles implementing step (c) of Figure \ref{fig:high-level-overview}, ``cleaning up'' by removing the output $f(x)$ from the scaffold, computation, and $x$ tiles.
    Though not shown, negative strength interactions are necessary between the second-to-top row of computation tiles and some of the right-growing cleanup tiles, to ensure that the right end of the row is properly detected. That is, there are two types of cleanup tiles growing right, one to detach the interior tiles, and one to detach the final rightmost computation tile. Since the east-west bonds between cleanup tiles are greater than 1, the negative north-south glue strengths between interior cleanup tiles and the second-to-rightmost blue tile -- and between the rightmost cleanup tile and the interior computation tiles -- must be strength -2 to ensure that the second-to-rightmost blue tile cannot stably attach except where intended.

}

%\opt{normal}{\mainConstructionDetails}
\mainConstructionDetails

%TODO: perhaps add some more discussion of the details of Figures \ref{fig:computation}, \ref{fig:position-for-cleanup}, and \ref{fig:cleanup}. In particular, Lila could indicate what was not clear on a first or second reading and we should emphasize these points.

%TODO: for journal version, it is probably possible to extend the construction to 1) non-constant row function $f$ using cooperation (but is it really useful?) 2) single-tile seed 3) cleaning junk

%% file: 5_conclusion.tex
\section{Conclusion}
\label{sec-conclusion}

\opt{normal}{
    We have shown two main results in the aTAM with negative glue strengths, under the standard assumption of \emph{irreversible} attachment, meaning attachments that only occur with strength at least the temperature $\tau$. The first result is that the amount of tile reuse afforded by the ability to detach tiles with negative glue strengths is fundamentally limited. After $t$ steps of assembly, $\Omega(t)$ tiles are permanently bound, unable to detach via negative glue strengths, and can only be detached by supplying external energy. The second result is a positive result that attempts to make do with this limitation: an $s(n)$-space-bounded Turing machine may be simulated for arbitrarily many steps, while ensuring that no intermediate assembly grows larger than $O(s(n))$.
}

    \newcommand{\appE}{Space-bounded ``computation'' as an end goal is not the only application of negative glue strengths, of course. Doty, Lutz, Patitz, Summers, and Woods \cite{RNSSA} study the problem of generating uniform random distributions on the finite sets using the independent flips of a fair coin afforded by the random selection of competing tile types in the aTAM (a non-trivial problem when the cardinality of the set is not a power of the number of competing tile types), and find a tradeoff between the closeness to uniformity of the distribution obtained and the space required for sampling. They exhibit a construction imposing a perfectly uniformly distribution on the set $\{0,1,\ldots,n-1\}$ that assembles a structure of width $\floor{\log n} + 1$ and \emph{expected} height at most 2, essentially implementing von Neumann's \emph{rejection method} of flipping $\floor{\log n} + 1$ fair coins repeatedly and stopping the first time that they encode a number smaller than $n$. It is very unlikely (probability at most $2^{-20}$) to take more than 20 attempts. But using this method in a construction such as that of \cite{USA},\footnote{The main construction of \cite{USA} shows how a ``universal'' tile set can be constructed that can be ``programmed'' through appropriate selection of a seed assembly to simulate the growth of any tile assembly system in a wide class of systems termed ``locally consistent'' (see \cite{USA} for details). In this discussion, we are concerned only with the fact that the construction of \cite{USA} 1) requires random numbers to be generated in a bounded space at many points throughout assembly, and 2) would be improved if the distribution of these numbers were perfectly uniform instead of ``close to uniform'' as in \cite{USA}.} in which many (perhaps more than $2^{20}$) copies of this experiment repeat throughout assembly, could increase the likelihood of growing too high. Even a single occurrence of a too-high subassembly will destroy the entire construction. Though we omit the details in the extended abstract, it is straightforward to augment the construction of \cite{USA} (which uses a variant of the random number selector of \cite{RNSSA}) with negative glue strengths to implement perfectly uniform selection of random numbers, thus improving the fidelity of the simulation of \cite{USA}, while providing an absolute guarantee on the space bound.
}
\opt{normal}{\appE}

\enlargethispage{1ex} % I know, it's bad to cheat...

%The construction we propose in  Section~\ref{sec-tm} can be adapted to other variants of the models (as described in Appendix~\ref{subsec-model-choices}).

There are other uses of negative glues in the aTAM.
For instance, we are able to improve the best known tile complexity (number of tile types) required to uniquely assemble a ``thin'' rectangle, i.e., an $n \times k$ rectangle with $k < \log n / (\log \log n - \log \log \log n)$. In the standard aTAM the tile complexity of this shape is known to be
%$\Omega \left(\frac{n^{1/k}}{k} \right)$~\cite{AGKS05}.
$\Omega (\frac{n^{1/k}}{k} )$~\cite{AGKS05}.
With the model of negative glue strengths we are able to improve this to $O\left( \sqrt{\log n} \right)$, %. Curiously, it is easier to create a \emph{thick} rectangle than a thin rectangle; we are able to
by first building a thick rectangle and using negative glues to ``cut out'' a thinner rectangle of the same length. % We omit the construction in this extended abstract.
\opt{dnasubmission}{We refer to Section~\ref{sec:other-uses} for more applications.}

Other questions related to this work include the experimental aspects of such a model, for example, how repulsive forces can be realized on DNA tiles, and how to ``clean'' and ``recycle'' the junk introduced during the assembly.

%% file: 6_appendix.tex
\section{Appendix}
\label{sec-appendix}

\modelDiscussionSection

\subsection{Definitions of Other Models}

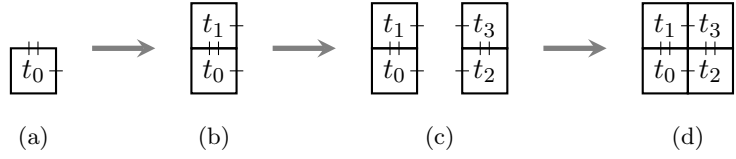
\begin{figure}[!ht]
      \begin{center}
        \input{figures/2handed.tex}
      \end{center}
      \caption{\figuresize Typical example of two-handed assembly, at temperature $\tau = 2$. The segments between tiles represent the bonds, the number of segments encodes the strength of the bond (here,~$1$ or~$2$). In the seeded, single tile model with seed $\sigma = t_0$, the assembly at step (b) would be terminal.}
      \label{fig:2handed}
    \end{figure}

    To define unseeded assembly, it suffices to drop $\sigma$ from the definition of TAS, and define the base case of a producible assembly as any individual tile. To define \emph{two-handed} assembly (a.k.a., \emph{multiple tile model}~\cite{AGKS05}), it suffices to change the first recursive case to state that legal attachment events are between any two producible assemblies, such that they can be positioned in such a way that the cut separating them has strength $\geq \tau$ (i.e., can be stably attached). Then, an assembly $\alpha$ is \emph{terminal} if for every producible assembly $\beta$, $\alpha$ and $\beta$ cannot be stably attached. 
    Figure~\ref{fig:2handed} illustrates the new behaviors allowed by the two-handed variant.

\subsection{Proof of Theorem \ref{thm-impossibility}}\label{sec:proof-impossibility}

\paragraph{Theorem \ref{thm-impossibility}.} Let $\calT$ be a TAS, and let $S$ be a multiset of assemblies producible by $\calT$ after $t\in\N$ steps. Then $\Phi(S) \geq t / 2$.

\impossibilityProof

The proof works for two-handed systems, and since single-tile addition systems are simply two-handed systems with an extra constraint on legal attachment operations, the proof applies to single-tile addition systems as well. The proof also works both for seeded and unseeded systems. As there is no property of the 2D plane or grid graphs employed in our proof, the proof applies to the \emph{irreversible} version of the graph-based self-assembly model studied in a reversible context by Reif, Sahu, and Yin \cite{ReifSahYin05}.

%\amortizedDiscussion

%\otherModelDefns

%\modelDiscussion
%\subsection{Definitions of Other Models}
%The definition of the model given in Section \ref{sec-tam-informal} can be easily modified to allow for other possibilities discussed in Section \ref{subsec-model-choices}.
%
%\otherModelDefns

%\subsection{Details of Main Construction}
%\label{subsec-main-construction-details}
%This section describes in more detail the construction shown at a high-level in Figure \ref{fig:high-level-overview}.

\subsection{Formal Definition of Constant-Row Computability}\label{sec:def-CRC}
The following is a formal definition of the class of functions that we call constant-row computable. The definition employs the standard model of nonnegative glue strengths, in which strengths are only positive between equal glues, so the glues themselves have both a ``type'' and a strength.

\constantRowDefn

%\mainConstructionDetails

\subsection{Other Uses of Negative Glue Strengths}\label{sec:other-uses}
\appE

%% file: niisa.bbl
\begin{thebibliography}{10}

\bibitem{AGKS05}
Gagan Aggarwal, Qi~Cheng, Michael~H. Goldwasser, Ming-Yang Kao, Pablo~Moisset
  de~Espan\'{e}s, and Robert~T. Schweller.
\newblock Complexities for generalized models of self-assembly.
\newblock {\em SIAM Journal on Computing}, 34:1493--1515, 2005.

\bibitem{CLRS01}
T.~H. Cormen, C.~E. Leiserson, R.~L. Rivest, and C.~Stein.
\newblock {\em Introduction to Algorithms}.
\newblock MIT Press, 2001.

\bibitem{USA}
David Doty, Jack~H. Lutz, Matthew~J. Patitz, Scott~M. Summers, and Damien
  Woods.
\newblock Intrinsic universality in self-assembly.
\newblock In {\em Proceedings of the 27th International Symposium on
  Theoretical Aspects of Computer Science}, 2009.
\newblock to appear.

\bibitem{RNSSA}
David Doty, Jack~H. Lutz, Matthew~J. Patitz, Scott~M. Summers, and Damien
  Woods.
\newblock Random number selection in self-assembly.
\newblock In {\em Proceedings of The Eighth International Conference on
  Unconventional Computation (Porta Delgada (Azores), Portugal, September 7-11,
  2009)}, 2009.

\bibitem{ReifSahYin05}
John~H. Reif, Sudheer Sahu, and Peng Yin.
\newblock Complexity of graph self-assembly in accretive systems and
  self-destructible systems.
\newblock In {\em DNA}, pages 257--274, 2005.

\bibitem{RotWin00}
Paul W.~K. Rothemund and Erik Winfree.
\newblock The program-size complexity of self-assembled squares (extended
  abstract).
\newblock In {\em STOC '00: Proceedings of the thirty-second annual ACM
  Symposium on Theory of Computing}, pages 459--468, New York, NY, USA, 2000.
  ACM.

\bibitem{SchYau79}
Richard Schoen and Shing-Tung Yau.
\newblock On the positive mass conjecture in general relativity.
\newblock {\em Communications in Mathematical Physics}, 65(45):45--76, 1979.

\bibitem{Seem82}
Nadrian~C. Seeman.
\newblock Nucleic-acid junctions and lattices.
\newblock {\em Journal of Theoretical Biology}, 99:237--247, 1982.

\bibitem{SolCooWinBru08}
David Soloveichik, Matthew Cook, Erik Winfree, and Jehoshua Bruck.
\newblock Computation with finite stochastic chemical reaction networks.
\newblock {\em Natural Computing}, 7(4):615--633, 2008.

\bibitem{Wang61}
Hao Wang.
\newblock Proving theorems by pattern recognition -- {II}.
\newblock {\em The Bell System Technical Journal}, XL(1):1--41, 1961.

\bibitem{Winf98}
Erik Winfree.
\newblock {\em Algorithmic Self-Assembly of {D}{N}{A}}.
\newblock PhD thesis, California Institute of Technology, June 1998.

\end{thebibliography}
